\documentstyle[12pt,chibaep]{article}
\def\@maketitle{\newpage
 \null
 {\normalsize \tt \begin{flushright}
  \begin{tabular}[t]{l} \@date
  \end{tabular}
 \end{flushright}}
 \begin{center}
 \vskip 2em
 {\LARGE \@title \par} \vskip 1.5em {\large \lineskip .5em
 \begin{tabular}[t]{c}\@author
 \end{tabular}\par}
 \end{center}
 \par
 \vskip 1.5em}
\def\underset#1\to#2{\mathop{#2}\limits_{#1}}
\def\Ge2{\Ge^2}
\def\GL2{\GL^2}
\def\Ga{\alpha}

\def\Gc{\chi}
\def\Gd{\delta}
\def\Ge{\epsilon}
\def\Gf{\phi}
\def\Gg{\gamma}
\def\Gh{\eta}

\def\Gj{\varphi}

\def\Gl{\lambda}
\def\Gm{\mu}
\def\Gn{\nu}
\def\Gp{\pi}
\def\Gq{\theta}

\def\Gs{\sigma}
\def\Gt{\tau}

\def\Gx{\xi}
\def\Gy{\psi}

\def\GF{\Phi}

\def\GL{\Lambda}
\def\GP{\Pi}

\def\GS{\Sigma}
\def\GW{\Omega}

\def\pd{\partial}
\def\tr{{\rm tr}}

\def\slala{\not \hskip-2pt}

\newcommand{\eq}[1]{eq.(\ref{#1})}

\title{
Gauge Independent Phase Structure of Gauged Nambu-Jona-Lasinio
and Yukawa Models
}
\author{
  {\sc Kei-ichi Kondo} \thanks{
    e-mail address: kondo@tansei.cc.u-tokyo.ac.jp;
                    kondo@cuphd.nd.chiba-u.ac.jp}
  \vspace{0.5cm} \\
  {\it Department of Physics, Faculty of Science,}\\
and  \\
  {\it Graduate School of Science and Technology,
      \thanks{mailing address}} \\
  {\it Chiba University,} \\
  {\it Inage-ku, Chiba 263, Japan}
  \vspace{1.2em} \\
}
\begin{document}
\date{
  hep-th/9305186 \\
  CHIBA-EP-69 \\
  May  1993
}
\maketitle
\begin{abstract}
We investigate the critical behavior of the gauged NJL model
(QED plus 4-fermion interaction) and the gauged Yukawa model
by use of the inversion method.
  By calculating the gauge-invariant chiral condensate in the
inversion method to the lowest order, we derive the critical line
which separates the spontaneous chiral-symmetry breaking phase from
the chiral symmetric one. The critical exponent for the chiral order
parameter associated with the second order chiral phase transition is
shown to take the mean-field value together with possible logarithmic
correction to the mean-field prediction.
  All the above results are gauge-parameter independent and are
compared with the previous results obtained from the Schwinger-Dyson
equation for the fermion propagator.
\end{abstract}
\newpage
\section{Introduction}
The dynamical symmetry breaking due to the four-fermion interaction
has been applied recently to the top quark condensation scenario of the
electroweak symmetry breaking \cite{MTY89,Yamawaki91}.
\par
As the next step, an attractive
force due to heavy spinless boson exchange through Yukawa interaction
has been examined to consider the origin of the
four-fermion interaction and to explain the masses of fermions
other than the top quark by the top condensation alone.
In this scenario, top quark condensate is mainly not due to the
vacuum expectation value of the spinless boson but to the attractive
force of the strong Yukawa coupling  \cite{KTY90}.
In the previous work  \cite{KTY90},
Tanabashi, Yamawaki and the author have analyzed this
possibility in the framework of the Schwinger-Dyson (SD) equation
\cite{KTY90} and have shown that there exists a chiral phase transition
$\langle \bar \Gy \Gy \rangle \not=0$ even in the
$\langle  \Gs \rangle =0$ phase in the strong Yukawa coupling region.
\par
However, this method of analysis has the following drawbacks.
We notice that the pseudoscalar Yukawa interaction
mediated by $\Gp$ gives attractive force but the
Yukawa interaction by $\Gs$ gives repulsive force.
Therefore, $U(1)_L\times U(1)_R$ symmetric Yukawa
model can not have the spontaneous chiral symmetry
breaking ($\Gc$SB) solution in the  $\langle  \Gs
\rangle =0$ phase because of the absence
(cancellation) of attractive force
\cite{KTY90,Tanabashi91}. This contradicts with the
result of lattice Monte Carlo simulation
\cite{HN89}. \par Another disadvantage is common to
the SD equation approach \cite{Kondo92p}. The SD
equation approach is restricted to the special gauge
and the gauge invariance of the result is not
necessarily guaranteed. This is because the fermion
self-energy treated in the SD equation is not gauge
invariant by nature. \par To resolve these
difficulties, we use the inversion method developed
recently by Fukuda and his collaborators
\cite{Fukuda88,UKF90}. The inversion method is the
extension of the usual effective action formalism
and the inversion process corresponds to the
Legendre transformation. By introducing an
appropriate source term $J {\cal O}$ into the
lagrangian, the expectation value of the (elementary
or composite) operator ${\cal O}$ is calculated in
the same procedure as  the perturbation theory. The
expectation value  $\langle {\cal O} \rangle$
vanishes for all orders of the perturbation in the
coupling constant $g$ if $J=0$. However, by
inverting  the function $\Gf(J) = \langle {\cal O}
\rangle$   of the source $J$  looking at $\Gf$ as an
quantity of order unity, the non-perturbative
solution is obtained.
 From this, the effective potential $V(\Gf)$ for the order
parameter $\langle {\cal O} \rangle$ is obtained by integration
through $J=\pd V(\Gf)/\pd \Gf$.
Moreover the inversion method can be improved systematically. Thus the
inversion method can be used to determine the  phase structure including
the order of the transition, the critical coupling and the critical
exponent. Actually the inversion method was successfully applied to
strong coupling phase of QED \cite{UKF90}.
\par
In this paper we choose the gauge-invariant order parameter
$\Gf = \langle \bar \Gy \Gy \rangle$
and apply the inversion method to the gauged Nambu-Jona-Lasinio
(NJL \cite{NJL61})
model and the gauged Yukawa model.
In the course of the calculation, gauge-parameter-independence is
explicitly elucidated and hence the gauge invariance is obvious.
We show that the dynamical $\Gc$SB occurs at strong Yukawa coupling,
i.e., there exists the $\langle \bar \Gy \Gy \rangle \not=0$ and
$\langle  \Gs \rangle =0$ phase, even in the $U(1)_L\times U(1)_R$
symmetric Yukawa model.

\newpage
\section{Gauged NJL model}
We consider the following gauged NJL \cite{NJL61} four-fermion model
distinguished by the group $G$:
\begin{equation}
 {\cal L}_0
 = \bar \Gy (i \slala \pd -m_0) \Gy - {1 \over 4} F_{\Gm \Gn}F^{\Gm \Gn}
   - e \bar \Gy \slala A \Gy
   + {g_F^2 \over 2}
   \left[ (\bar \Gy {\bf 1} \Gy)^2+(\bar \Gy i \Gg_5 T_a \Gy)^2
\right] .
\end{equation}
Here ${\bf 1}$ is the $N \times N$ unit matrix where
$N=1$ for $G=Z(2), U(1)$ and $N=2$ for $G=SU(2)$,
and $T_a$ ($a=1,2,...,d_G:=dim G$) is defined as
$T_a=0, d_G=0$ for $G=Z(2)$; $T_a=1, d_G=1$ for $G=U(1)$;
$T_a=\Gt^a$ (Pauli matrices), $d_G=3$  for $G=SU(2)$.
\par
We start from the lagrangian ${\cal L}={\cal L}_0+{\cal L}_{GF}$
involving the gauge fixing term
${\cal L}_{GF}=-{1 \over 2\Gx} (\pd_\Gm A^\Gm)^2$ with gauge fixing
parameter $\Gx$:
\begin{eqnarray}
 {\cal L} &=& \bar \Gy (i \slala \pd -m_0) \Gy
 - {1 \over 2} A^{\Gm} [- g_{\Gm \Gn} \pd^2
                         + (1-\Gx^{-1})\pd_\Gm \pd_\Gn] A^{\Gn}
\nonumber\\
 &&- e \bar \Gy \slala A \Gy
 + {g_F^2 \over 2} \left[ (\bar \Gy {\bf 1} \Gy)^2+(\bar \Gy i \Gg_5 T_a \Gy)^2
\right] .
\end{eqnarray}
By introducing the auxiliary field $\Gs,\Gp_a$, the lagrangian
\footnote{
For $G=U(1)$, this lagrangian is invariant under the (continuous) chiral
transformation
$$
\bar \Gy \rightarrow \bar \Gy  e^{i \Gg_5 \Gq}, \
 \Gy \rightarrow  e^{i \Gg_5 \Gq}  \Gy , \
 \Gs \rightarrow  \Gs \cos 2\Gq + \Gp \sin 2\Gq, \
 \Gp \rightarrow  -\Gs \sin 2\Gq + \Gp \cos 2\Gq .
$$
The lagrangian is invariant also under the discrete chiral
transformation
$
\bar \Gy \rightarrow \bar \Gy  i \Gg_5, \
 \Gy \rightarrow  i \Gg_5  \Gy , \
 \Gs \rightarrow - \Gs, \
 \Gp \rightarrow - \Gp .
$
The four-fermion model with either
 $(\bar \Gy \Gy)^2$ or $(\bar \Gy i \Gg_5 \Gy)^2$ is invariant only
under the discrete transformation.
}
is equivalently rewritten as
\begin{eqnarray}
 {\cal L} = \bar \Gy (i \slala \pd ) \Gy
 - {1 \over 2} A^{\Gm} [- g_{\Gm \Gn} \pd^2
                        + (1-\Gx^{-1})\pd_\Gm \pd_\Gn ] A^{\Gn}
 + {1 \over 2 } (\Gs^2 + \Gp_a^2)
\nonumber\\
 - e \bar \Gy \slala A \Gy
 - g_F \bar \Gy (\Gs {\bf 1} + i\Gg_5 T_a\Gp_a) \Gy + {m_0 \over g_F} \Gs .
\end{eqnarray}

\par
We introduce the (bilocal) source term to derive the
generating functional as follows:
\begin{eqnarray}
S_J = - \int d^4x d^4y [J(x,y) \bar \Gy(x) \Gy(y)
+ {1 \over 2} A^{\Gm}(x)J_{\Gm \Gn}^A(x,y)A^{\Gn}(y)] .
\end{eqnarray}
We decompose the total action $S_{total}=S+S_J$ into
\begin{eqnarray}
 S_{total} &=& S_0^J + S_I,
\nonumber\\
 S_0^J &=& \int d^4x d^4y \Biggr[
 \bar \Gy(x) (i \slala \pd \Gd^4(x,y) + J(x,y)) \Gy(y)
\nonumber\\
&&
- {1 \over 2} A^{\Gm}(x) \{ [- g_{\Gm \Gn} \pd^2
                        + (1-\Gx^{-1})\pd_\Gm \pd_\Gn] \Gd^4(x,y)
                        -J_{\Gm \Gn}^A(x,y) \} A^{\Gn}(y)
\nonumber\\
&&
 + {1 \over 2 } (\Gs^2(x) + \Gp_a^2(x))\Gd^4(x,y)
 +  {m_0 \over g} \Gs(x)\Gd^4(x,y)
  \Biggr],
\nonumber\\
 S_I &=& - e \int d^4x\bar \Gy(x) \slala A(x) \Gy(x)
 - g_F \int d^4x \bar \Gy(x) (\Gs(x) {\bf 1} + i\Gg_5 T_a \Gp_a(x)) \Gy(x) .
\end{eqnarray}

Now the generating functional $W[J,J^A]$ is defined as
\begin{eqnarray}
 W[J,J^A]  &:=&   \ln \int [d\bar\Gy][d\Gy][dA][d\Gs][d\Gp]
 \exp (-S_{total})
\nonumber\\
&=& \ln \int [d\bar\Gy][d\Gy][dA][d\Gs][d\Gp] \exp (-S_0^J)
 + \ln \langle \exp (-S_I) \rangle,
\end{eqnarray}
where
\begin{eqnarray}
\langle F \rangle :=
{\int [d\bar\Gy][d\Gy][dA][d\Gs][d\Gp] \exp (-S_0^J) F
\over \int [d\bar\Gy][d\Gy][dA][d\Gs][d\Gp] \exp (-S_0^J)}.
\end{eqnarray}
The second part has the following cumulant expansion:
\begin{eqnarray}
\ln \langle \exp (-S_I) \rangle
= \ln \sum_{n=0}^\infty {1 \over n!} \langle (-S_I)^n \rangle
=  \langle -S_I \rangle + {1 \over 2}[\langle (-S_I)^2 \rangle
-\langle -S_I \rangle^2] + \cdots .
\end{eqnarray}
\par
When $\langle A^\Gm(x) \rangle = 0 = \langle \Gp_a(x) \rangle$,
we obtain
\begin{eqnarray}
\langle -S_I \rangle  = g_F \int d^4 x
\langle \Gs(x) \rangle \langle \bar \Gy(x) {\bf 1} \Gy(x) \rangle,
\end{eqnarray}
and
\begin{eqnarray}
\langle (-S_I)^2 \rangle  &=& \int d^4 x \int d^4 y
\Biggr\{ e^2 \langle A^\Gm(x) A^\Gn(y)  \rangle
\langle \bar \Gy(x)\Gg_\Gm \Gy(x)
        \bar \Gy(y)\Gg_\Gn \Gy(y)  \rangle
\nonumber\\
&& + g_F^2 \Big[ \langle \Gs(x) \Gs(y)  \rangle
\langle \bar \Gy(x) {\bf 1} \Gy(x)
        \bar \Gy(y) {\bf 1} \Gy(y)  \rangle
\nonumber\\
&&  + \langle \Gp_a(x) \Gp_b(y)  \rangle
\langle \bar \Gy(x) i \Gg_5 T_a \Gy(x)
        \bar \Gy(y) i \Gg_5 T_b \Gy(y)  \rangle \Big] \Biggr\} .
\end{eqnarray}

For the uniform source
$J(x,y) = J\Gd^4(x-y)$ and $J_{\Gm \Gn}^A(x,y)=J_{\Gm \Gn}^A\Gd^4(x-y)$,
the first part of the generating functional corresponding to $S_0$
in the absence of fermion bare mass $m_0=0$ reads
\begin{eqnarray}
 W_0[J,J^A]
&=&  \GW \int {d^4p \over (2\Gp)^4} \tr [\ln (S^{-1}(p) +J)]
 - {1 \over 2} \GW \int {d^4k \over (2\Gp)^4}
 \tr [\ln (D_{\Gm \Gn}^{-1}(k) +J_{\Gm \Gn}^A)]
\nonumber\\
 && - {1 \over 2}  \GW \int {d^4k \over (2\Gp)^4}
 \{ \tr [\ln G_{\Gs}^{-1}(k)]+ \tr [\ln G_{\Gp}^{-1}(k)] \},
\end{eqnarray}
where $\GW$ is the space-time volume and  the respective inverse
propagator is defined by
\begin{eqnarray}
S^{-1}(p) &=& \slala p,
\nonumber\\
D_{\Gm \Gn}^{-1}(k) &=& g_{\Gm \Gn}k^2-(1-\Gx^{-1})k_\Gm k_\Gn,
\nonumber\\
G_\Gs^{-1}(k) &=& G_\Gp^{-1}(k) = 1.
\end{eqnarray}

The remaining second part of the generating functional is written as
\begin{eqnarray}
 W_I[J,J^A]/\GW
 &=&  {e^2 N N_f \over 2}
 \int {d^4k \over (2\Gp)^4} D^{\Gm \Gn}(k) \GP_{\Gm \Gn}(k)
\nonumber\\
&& + {g_F^2 N N_f \over 2} \int {d^4k \over (2\Gp)^4}
 [G_\Gs(k) \GP_\Gs(k)+ d_G G_\Gp(k) \GP_\Gp(k)]
\nonumber\\
&&+g_F \langle \Gs \rangle
 \langle \bar \Gy \Gy \rangle
 - {1 \over 2}  g_F^2 \langle \Gs \rangle^2
 \langle \bar \Gy \Gy \rangle^2 + \cdots ,
\end{eqnarray}
where $N_f$ denotes the species of the fermion and the vacuum
polarization function for each fermion is given by
\begin{eqnarray}
 \GP_{\Gm \Gn}(k) &=& \int {d^4p \over (2\Gp)^4}
 \tr \left[S_J(p) \Gg_\Gm S_J(p+k) \Gg_\Gn \right],
\nonumber\\
 \GP_{\Gs}(k) &=& \int {d^4p \over (2\Gp)^4}
 \tr \left[S_J(p)  S_J(p+k) \right],
\nonumber\\
 \GP_{\Gp}(k) &=& \int {d^4p \over (2\Gp)^4}
 \tr \left[S_J(p) i\Gg_5 S_J(p+k) i\Gg_5  \right],
\end{eqnarray}
with the fermion propagator $S_J(p)$ with source term $J$:
\begin{eqnarray}
S_J^{-1}(p) = S^{-1}(p)+J = \slala p +J .
\end{eqnarray}
\par
Therefore we obtain the generating functional to  the
order $\Ga,G:=g_F^2$:
\begin{eqnarray}
 W[J,J^A]/\GW
 &=&
 N N_f \int {d^4p \over (2\Gp)^4} \tr [\ln S_J^{-1}(p)]
 - {1 \over 2}  \int {d^4k \over (2\Gp)^4}
 \tr [\ln (D_{\Gm \Gn}^{-1}(k) +J_{\Gm \Gn}^A)]
\nonumber\\
 && - {1 \over 2}  \int {d^4k \over (2\Gp)^4}
 \{ \tr [\ln G_{\Gs}^{-1}(k)]+ \tr [\ln G_{\Gp}^{-1}(k)] \}
\nonumber\\
 && + g_F \langle \Gs \rangle   \langle \bar \Gy \Gy \rangle
+ {g_F^2 N N_f \over 2} \int {d^4k \over (2\Gp)^4}
 [G_\Gs(k) \GP_\Gs(k)+ d_G G_\Gp(k) \GP_\Gp(k)]
 \nonumber\\
&&+ {e^2 N N_f \over 2}
 \int {d^4k \over (2\Gp)^4} D^{\Gm \Gn}(k) \GP_{\Gm \Gn}(k) .
\end{eqnarray}

The chiral condensation is obtained from the generating functional as
\begin{eqnarray}
\Gf(J) &:=& \langle \bar \Gy \Gy \rangle_J
 = {1 \over \GW} {\pd W[J,J^A] \over \pd J}\Big|_{J^A=0} .
\end{eqnarray}
Namely,
\begin{eqnarray}
 \langle \bar \Gy \Gy \rangle_J
 &=&  {\pd \over \pd J}
 N N_f \int {d^4p \over (2\Gp)^4} \tr [\ln S_J^{-1}(p) ]
 + g_F
 \langle \Gs \rangle
 {\pd \over \pd J} \langle \bar \Gy \Gy \rangle \Big|_{J^A=0}
\nonumber\\
 &&+ {g_F^2 N N_f \over 2} {\pd \over \pd J}
\int {d^4k \over (2\Gp)^4}
 [G_\Gs(k) \GP_\Gs(k)+d_G G_\Gp(k) \GP_\Gp(k)]\Big|_{J^A=0}
 \nonumber\\
&&+ {e^2 N N_f \over 2} {\pd \over \pd J}
 \int {d^4k \over (2\Gp)^4} D^{\Gm \Gn}(k) \GP_{\Gm \Gn}(k)\Big|_{J^A=0} ,
\label{order}
\end{eqnarray}
where the right-hand-side of this equation should be understood to be
calculated at $J^A=0$.
The vacuum expectation value of the auxiliary field
is related to the chiral condensation as
\begin{eqnarray}
\langle \Gs \rangle =
 g_F \langle \bar \Gy \Gy \rangle - {m_0 \over g_F}.
\end{eqnarray}

Introducing the fermion momentum cutoff $\GL_f$, we obtain
\begin{eqnarray}
{\langle \bar \Gy \Gy \rangle \over N N_f}
&=& - \int^{\GL_f} {d^4p \over (2\Gp)^4} \tr {1 \over \slala p +J}
= - {\pd \over \pd J}
\int {d^4p \over (2\Gp)^4} \tr [\ln S_J^{-1}(p)]
\nonumber\\
&=& {\GL_f^2 \over 4\Gp^2} J + J^3{1 \over 4\Gp^2}
\ln \left({\GL_f^2 \over J^2}-1\right),
\end{eqnarray}
which yields
\begin{eqnarray}
 g_F \langle \Gs \rangle {\pd \over \pd J}
 \langle \bar \Gy \Gy \rangle
 &=&  {g_F^2 \over 2} {\pd \over \pd J}
 \langle \bar \Gy \Gy \rangle^2
\nonumber\\
&=&   g_F^2
\left( {N N_f\GL_f^2 \over 4\Gp^2} \right)^2
\left[ J + 4 {J^3 \over \GL_f^2}
\ln \left({\GL_f^2 \over J^2}-1\right) \right]
+ {\cal O}(J^5) .
\end{eqnarray}
It is observed \cite{UKF90}
that \eq{order} gives gauge-parameter independent chiral condensates,
since
\begin{eqnarray}
 D^{\Gm \Gn}(k) \GP_{\Gm \Gn}(k)
= {1 \over k^2} \left[ g^{\Gm \Gn}-(1-\Gx){k^{\Gm}k^{\Gn} \over k^2}
\right] (g_{\Gm \Gn}k^2-k_{\Gm}k_{\Gn}) \GP(k^2)
=  3\GP(k^2).
\end{eqnarray}
Gauge-invariant
Pauli-Villars regularization \cite{IZ} gives
\begin{eqnarray}
 \GP(k^2) &=& - {\Ga \over 3\Gp} \Biggr\{- \ln {\GL^2 \over J^2}
 + 2 \left(1+{2J^2 \over k^2}\right)
 \left[\left({4J^2 \over k^2}-1\right)^{1/2}
 \cot^{-1}\left({4J^2 \over k^2}-1\right)^{1/2} -1 \right]
\nonumber\\
&& - 2\left(1+{2\GL^2 \over k^2}\right)^{1/2}
 \left[\left({4\GL^2 \over k^2}-1\right)^{1/2}
 \cot^{-1}\left({4\GL^2 \over k^2}-1\right)^{1/2} -1 \right] \Biggr\},
\end{eqnarray}
with $\GL$ being the mass of the auxiliary field.
After taking $\pd /\pd J$, the dependence on $\GL$ disappear as
\begin{eqnarray}
 {\pd \over \pd J} \GP(k^2)
 &=& - {\Ga \over 3\Gp} {8J \over k^2}
 \Biggr\{  -{3 \over 2}
+ \left({4J^2 \over k^2}-1\right)^{1/2}
 \cot^{-1}\left({4J^2 \over k^2}-1\right)^{1/2}
\nonumber\\
 &&+\left(1+{2J^2 \over k^2}\right)
 \left({4J^2 \over k^2}-1\right)^{-1/2}
 \cot^{-1}\left({4J^2 \over k^2}-1\right)^{1/2}\Biggr\}.
\end{eqnarray}
Hence we obtain
\begin{eqnarray}
 {\pd \over \pd J} {e^2 \over 2}
 \int^{\GL_p} {d^4k \over (2\Gp)^4} D^{\Gm \Gn}(k) \GP_{\Gm \Gn}(k)
= {3\GL_p^2 \over 8\Gp^2} \Ga J - \Ga J^3{3 \over 8\Gp^3}
\ln^2 {J^2 \over \GL_p^2} + {\cal O}(J^3 \ln J^2),
\end{eqnarray}
where $\GL_p$ denotes the photon momentum cutoff.
On the other hand, the vacuum polarization effect of the auxiliary field
is calculated as follows.
We first notice that
\begin{eqnarray}
\int {d^4k \over (2\Gp)^4}
 [G_\Gs(k) \GP_\Gs(k)]\Big|_{J^A=0}
&=& \int {d^4p \over (2\Gp)^4} \int {d^4q \over (2\Gp)^4}
 G_\Gs(p-q) \tr[S(p)S(q)]
\nonumber\\
&=& \int {d^4p \over (2\Gp)^4} \int {d^4q \over (2\Gp)^4}
  {4(p \cdot q + J^2)G_\Gs(p-q) \over (p^2-J^2)(q^2-J^2)},
\end{eqnarray}
and
\begin{eqnarray}
\int {d^4k \over (2\Gp)^4}
 [G_\Gp(k) \GP_\Gp(k)]\Big|_{J^A=0}
&=& \int {d^4p \over (2\Gp)^4} \int {d^4q \over (2\Gp)^4}
 G_\Gp(p-q) \tr[i\Gg_5 S(p)i\Gg_5 S(q)]
\nonumber\\
&=& \int {d^4p \over (2\Gp)^4} \int {d^4q \over (2\Gp)^4}
  {4(p \cdot q - J^2)G_\Gp(p-q) \over (p^2-J^2)(q^2-J^2)}.
\end{eqnarray}
Summing up the above two terms, we obtain
\begin{eqnarray}
&& \int {d^4k \over (2\Gp)^4}
[G_\Gs(k)\GP_\Gs(k)+d_G G_\Gp(k)\GP_\Gp(k)]
\nonumber\\
&=& \int {d^4p \over (2\Gp)^4} \int {d^4q \over (2\Gp)^4}
  {4(1+d_G) (p \cdot q) +4(1-d_G)J^2
  \over (p^2-J^2)(q^2-J^2)}
\nonumber\\
&=& {1-d_G \over 4} \left[  \int {d^4k \over (2\Gp)^4} {4J \over p^2-J^2}
\right]^2
\nonumber\\
&=&  {1-d_G \over 4} \left(
{\langle \bar \Gy \Gy \rangle \over N N_f} \right)^2 ,
\end{eqnarray}
where we have used the result of the angular integration:
$\int_0^\Gp d\Gq \sin^2\Gq \cos \Gq \equiv 0$.
Then this term identically vanishes for the $U(1)$ symmetric case
$d_G=1$. \footnote{This fact has been already pointed out in
\cite{KTY93}.}

\par
Therefore we obtain
\begin{eqnarray}
\Gf(J) &=&  {N N_f\GL_f^2 \over 4\Gp^2} J
\left[ 1+ \left(1+{1-d_G \over 4N N_f}\right) g
+ {\Ga\over \Ga_c} \right]
\nonumber\\
&&- {N N_f \over 4\Gp^2 \Gh}
J^3 \left[4\Gh g \ln {J^2 \over \GL_f^2}
 + {\Ga\over \Ga_c} \ln^2 {J^2 \over \GL_p^2} \right]
+ {\cal O}(J^3 \ln J^2),
\end{eqnarray}
where we have defined
\begin{eqnarray}
g :=  {N N_f g_F^2 \GL_f^2 \over 4\Gp^2},
\ \Ga_c := {2\Gp \over 3 \Gh}, \ \Gh := {\GL_p^2 \over \GL_f^2} .
\end{eqnarray}
Here $g$ is the dimensionless four-fermion coupling and
$\Ga_c$ is the critical coupling of pure QED obtained previously in
the inversion method by Ukita, Komachiya and Fukuda \cite{UKF90}.
The critical value $\Ga_c=2\Gp/3$ is in good agreement with the result
of the SD equation $\Ga_c=2.00$ for $\Gh=1$ and $N_f=1$ \cite{KN91}.
In the SD equation, $\Ga_c$ depends on the two cutoffs, but their
dependence is different from the result of the inversion method
\cite{KN91}. It should be remarked that the above
$\Ga_c$ has no $N_f$-dependence, which contradicts with the result of the
SD equation.  To derive $N_f$ dependence of $\Ga_c$ in the inversion
method, we need to take into account the next-to-leading order.

\par
Now we invert $\Gf(J) := \langle \bar \Gy \Gy
\rangle_J$  given above and get to the order $\Ga, g$:
\footnote{
For the coupling constant $\Gl$, $\Gf(J)$ is
calculated as
$\Gf(J)=f_0(J)+\Gl f_1(J)+\cdots = aJ+\Gl(bJ+cJ^3)+\cdots$.
Then the inversion yields
$J=h_0(\Gf)+\Gl h_1(\Gf)+\cdots
=A\Gf-\Gl(K\Gf-L\Gf^3)+\cdots$.
Looking $\Gf$ as order of unity implies the relation:
$A=a^{-1}, K=Aba^{-1}=ba^{-2}, L=A^3ca^{-1}=ca^{-4}.$
Therefore, putting $J=0$ gives the nontrivial solution
$\Gf= \sqrt{(\Gl K-A)/(L\Gl)}$ for $\Gl>\Gl_c:=A/K=a/b$,
besides the trivial one $\Gf=0$,
since the effective potential is given by
$V(\Gf)={1 \over 2}(A-\Gl K)\Gf^2+{\Gl \over 4} L \Gf^4$
from the relation $J=\pd V(\Gf)/\pd \Gf$.
Here $\Gl_c$ is interpreted as the second order phase transition point
when $L>0$ ($c>0$).

}
\begin{eqnarray}
J(\Gf)
&=& {4\Gp^2 \over N N_f\GL_f^2}
\left[ 1- \left(1+{1-d_G \over 4N N_f}\right)g
- {\Ga\over \Ga_c} \right] \Gf
\nonumber\\
&&+  {64 \Gp^6 \over \Gh N^3 N_f^3 \GL_f^8}
\left[ 4\Gh g \ln \left({16\Gp^4 \over N^2 N_f^2 \GL_f^6} \Gf^2 \right)
+ {\Ga \over \Ga_c}
\ln ^2 \left({16\Gp^4 \over N^2 N_f^2\GL_f^4 \GL_p^2} \Gf^2 \right)
\right] \Gf^3
\nonumber\\
&&+ {\cal O}(\Gf^3 \ln \Gf^2).
\label{inversion}
\end{eqnarray}
The effective potential $V(\Gf)$ of $\Gf$ is obtained from the relation
$J(\Gf)=dV(\Gf)/d\Gf$ as
\begin{eqnarray}
V(\Gf)
&=& {2\Gp^2 \over N N_f\GL_f^2} \left[ 1-
\left(1+{1-d_G \over 4N N_f}\right)g - {\Ga\over \Ga_c} \right]\Gf^2
\nonumber\\
&& +  {16 \Gp^6 \over \Gh N^3 N_f^3 \GL_f^8}
\left[ 4\Gh g \ln \left( {16\Gp^4 \over N^2 N_f^2 \GL_f^6} \Gf^2  \right)
+ {\Ga \over \Ga_c}
\ln ^2 \left({16\Gp^4 \over N^2 N_f^2\GL_f^4 \GL_p^2} \Gf^2
\right)  \right] \Gf^4
\nonumber\\
&&         + {\cal O}(\Gf^4 \ln \Gf^2) .
\end{eqnarray}
The sign of the first term determines the phase according to the
Landau-Ginzburg theory of the second order phase transition, since the
coefficient of $\Gf^4$ in the effective potential is positive at
least                for $\Gf/\GL^3 \ll 1$. Thus we obtain the
critical line of the gauged NJL model:
\begin{eqnarray}
 g = \left(1+{1-d_G \over 4N N_f}\right)^{-1}
 \left(1 - {\Ga\over \Ga_c}\right).
 \label{critical}
\end{eqnarray}
In particular,  for $U(1)$ symmetric case
\begin{eqnarray}
 g = 1 - {\Ga\over \Ga_c},
 \label{criticalline}
\end{eqnarray}
which should be compared with the critical line in the quenched
approximation \cite{KMY89} where the vacuum polarization to the
photon propagator is neglected:
\footnote{The critical line
\eq{quenchcriticalline} locates below \eq{criticalline}.
But \eq{quenchcriticalline} coincides with \eq{criticalline} to
the order ${\cal O}(\Ga)$ and has
the same slope as \eq{criticalline} at $\Ga=0$, since
$g_Q = 1-{\Ga \over \Ga_c}-{1 \over 4}({\Ga \over \Ga_c})^2 -
\cdots$.} \begin{eqnarray}
 g = g_Q := \cases{
 {1 \over 4} \left(1+\sqrt{1 - \Ga/\Ga_c^Q}\right)^2
 & ($\Ga<\Ga_c^Q$) \cr
 \le {1 \over 4} & ($\Ga=\Ga_c^Q$) \cr } ,
\label{quenchcriticalline}
\end{eqnarray}
with
\begin{eqnarray}
  \Ga_c^Q = {\Gp \over 3}.
\end{eqnarray}
Above the critical line \eq{criticalline},
the chiral symmetry is spontaneously broken.
In the region $0<\Ga<\Ga_c$, the critical line \eq{criticalline}
exhibits extremely good agreement with that obtained through the solution
(in the Landau gauge) of the SD equation for the fermion
propagator in the unquenched case, i.e., in the presence of the 1-loop
vacuum polarization  \cite{Kondo91}.
For super strong region $\Ga>4$, the critical line obtained from the
SD equation \cite{Gusynin90,Kondo91b,Rakow91}
deviates from the straight line \eq{criticalline}, since
\eq{criticalline} is obtained in the lowest order of the inversion
method.
\par
The chiral condensation behaves in the neighborhood of the critical line
as
\begin{eqnarray}
\langle \bar \Gy \Gy \rangle
\sim  {N N_f \sqrt{\Gh} \over 4\Gp^2} \GL_f^3
\Gt^{1/2} [\ln \Gt]^{-1/2},
\end{eqnarray}
for small derivation $\Gt$ from the critical line:
\begin{eqnarray}
\Gt := 1 - \left(1+{1-d_G \over 4N N_f}\right)g
- {\Ga \over \Ga_c}.
\end{eqnarray}

This indicates that the chiral order parameter takes the mean-field
critical exponent 1/2    in the neighborhood of the whole critical line
\eq{criticalline} as already shown by the analysis of SD equation in the
unquenched case \cite{Kondo91,Kondo91b}
and that there is a possibility of logarithmic correction to the
mean-field prediction in the gauged NJL model.
\par
All the above results are gauge-independent, while the SD result for the
gauged NJL model in the unquenched case is restricted to the Landau gauge
\cite{Kondo91,Kondo91b,Gusynin90,Rakow91}.
For strong coupling QED, it has been shown
that gauge-independent results in the framework of the SD equation are
obtained by taking into account the vertex correction even in the
unquenched case, i.e., in the presence of the 1-loop vacuum
polarization to the photon propagator \cite{KKM89,Kondo92}. In other
words, the result obtained in the Landau gauge is preserved in other
covariant gauges.

\newpage
\section{Gauged Yukawa model}
In this section we consider the gauged
\footnote{We assume that the scalar field $\Gs, \Gp$ is not
transformed under the gauge transformation
$
A_\Gm(x) \rightarrow A_\Gm(x)+\pd_\Gm \GL(x),
\bar \Gy(x) \rightarrow \bar \Gy(x) e^{ie\GL(x)},
\Gy(x) \rightarrow e^{-ie\GL(x)}\Gy(x).
$
To make the scalar part gauge-invariant, we should take
$
{1 \over 2}|(\pd_\Gm-ieA_\Gm)\Gj|^2+{\Gm^2 \over 2}|\Gj|^2
+{\Gl \over 4}(|\Gj|^2)^2
= {1 \over 2}[(\pd_\Gm \Gs)^2+(\pd_\Gm \Gp)^2]
+{\Gm^2 \over 2}(\Gs^2+\Gp^2)+{\Gl \over 4}(\Gs^2+\Gp^2)^2
-2e A^\Gm[\Gs \pd_\Gm \Gp-\Gp \pd_\Gm \Gs]+e^2 A_\Gm A^\Gm(\Gs^2+\Gp^2),
$
for the $U(1)$ scalar field $\Gj=\Gs+i\Gp$. However the
gauge-invariance   is not preserved for the naive Yukawa coupling
$
\bar \Gy (\Gs + i\Gg_5 \Gp)\Gy
=\bar \Gy_R \Gj \Gy_L+\bar \Gy_L \Gj^{\dagger} \Gy_R ,
$
unless Yukawa coupling is modified as in the standard electroweak
theory.  Such a case will be discussed elsewhere.
}
$G_L \times G_R$-symmetric Yukawa model with the lagrangian
\begin{eqnarray}
 {\cal L} &=& \bar \Gy (i \slala \pd  - m_0) \Gy
 - {1 \over 2} A^{\Gm} [- g_{\Gm \Gn} \pd^2
                        + (1-\Gx^{-1})\pd_\Gm \pd_\Gn ] A^{\Gn}
                        - {1 \over 2}\Gs(\pd^2-m_\Gs^2)\Gs
\nonumber\\
&&  - {1 \over 2}\Gp_a(\pd^2-m_\Gp^2)\Gp_a
 - e \bar \Gy \slala A \Gy
 - g_Y \bar \Gy (\Gs {\bf 1} + i\Gg_5 T_a \Gp_a) \Gy
 + {\Gl \over 4} (\Gs^2+\Gp_a^2)^2.
\end{eqnarray}
Here ${\bf 1}$ is the $N \times N$ unit matrix where
$N=1$ for $G=Z(2), U(1)$ and $N=2$ for $G=SU(2)$,
and $T_a$ ($a=1,2,...,d_G:=dim G$) is defined as
$T_a=0, d_G=0$ for $G=Z(2)$; $T_a=1, d_G=1$ for $G=U(1)$;
$T_a=\Gt^a$ (Pauli matrices), $d_G=3$  for $G=SU(2)$.
Actually the Yukawa interaction,
$
\bar \Gy (\Gs {\bf 1}+ i\Gg_5 T_a \Gp_a) \Gy
=\bar \Gy_R \GF \Gy_L+\bar \Gy_L \GF^{\dagger} \Gy_R ,
$
is $G_L \times G_R$-symmetric
for the scalar field $\GF=\Gs {\bf 1}+i T_a \Gp_a$.
\footnote{
When $m_\Gs \not= m_\Gp$, the lagrangian is invariant only under the
discrete chiral transformation,
while the lagrangian is invariant under the
continuous chiral transformation for $m_\Gs = m_\Gp$.
}
\par
To apply the inversion method for this lagrangian, we take
\begin{eqnarray}
 S_{total} &=& S_0^J + S_I,
\nonumber\\
 S_0^J &=& \int d^4x d^4y \Biggr[
 \bar \Gy(x) ([i \slala \pd -m_0]\Gd^4(x,y) + J(x,y)) \Gy(y)
\nonumber\\
&&
- {1 \over 2} A^{\Gm}(x) \{ [- g_{\Gm \Gn} \pd^2
                        + (1-\Gx^{-1})\pd_\Gm \pd_\Gn] \Gd^4(x,y)
                        -J_{\Gm \Gn}^A(x,y) \} A^{\Gn}(y)
\nonumber\\
&&
 - {1 \over 2 } \Gs(x)[\pd^2-m_\Gs^2]\Gd^4(x,y)\Gs(y)
 - {1 \over 2 } \Gp_a(x)[\pd^2-m_\Gp^2]\Gd^4(x,y)\Gp_a(y)
\nonumber\\
&&
  + {\Gl \over 4} (\Gs(x)^2+\Gp_a(x)^2)^2 \Gd^4(x,y)
  \Biggr],
\nonumber\\
 S_I &=& - e \int d^4x\bar \Gy(x) \slala A(x) \Gy(x)
 - g_Y \int d^4x \bar \Gy(x) [\Gs(x){\bf 1} + i\Gg_5 T_a \Gp_a(x)] \Gy(x) .
\end{eqnarray}

We follow the same steps as the gauged NJL model
assuming
$\langle A^\Gm(x) \rangle = 0 = \langle \Gp_a(x) \rangle$.
Then, replacing $g_F$ with $g_Y$ and
the auxiliary field propagator $G_\Gs(k), G_\Gp(k)$ with
the scalar field
propagators: \begin{eqnarray}
 G_\Gs(k) = (k^2+m_\Gs^2)^{-1}, \  G_\Gp(k) = (k^2+m_\Gp^2)^{-1},
\end{eqnarray}
we obtain
\begin{eqnarray}
 \langle \bar \Gy \Gy \rangle_J
 &=&  {\pd \over \pd J}
 N N_f \int {d^4p \over (2\Gp)^4} \tr [\ln S_J^{-1}(p) ]
+ g_Y
 \langle \Gs \rangle
 {\pd \over \pd J} \langle \bar \Gy \Gy \rangle \Big|_{J^A=0}
\nonumber\\
 &&+ {g_Y^2 N N_f \over 2} {\pd \over \pd J}
\int {d^4k \over (2\Gp)^4}
 [G_\Gs(k) \GP_\Gs(k)+ d_G G_\Gp(k) \GP_\Gp(k)]\Big|_{J^A=0}
 \nonumber\\
&&+ {e^2 N N_f \over 2} {\pd \over \pd J}
 \int {d^4k \over (2\Gp)^4} D^{\Gm \Gn}(k) \GP_{\Gm \Gn}(k)\Big|_{J^A=0}
{}.
\end{eqnarray}
We notice that
\begin{eqnarray}
&& \int {d^4k \over (2\Gp)^4}
 [G_\Gs(k) \GP_\Gs(k)+d_G G_\Gp(k) \GP_\Gp(k)]\Big|_{J^A=0}
\nonumber\\
&=& \int {d^4p \over (2\Gp)^4} \int {d^4q \over (2\Gp)^4}
  {4 \over (p^2-J^2)(q^2-J^2)}
  \left[
  {(p \cdot q + J^2) \over (p-q)^2+m_\Gs^2}
  +   d_G {(p \cdot q - J^2) \over (p-q)^2+m_\Gp^2} \right]
\nonumber\\
&=& \int {dp^2 \over 16\Gp^2} \int {dq^2 \over 8\Gp^3}
  {\Gp p^2 q^2 \over (p^2-J^2)(q^2-J^2)}
  [{\cal K}_A(p^2,q^2)-2J^2 {\cal K}_B(p^2,q^2)],
\end{eqnarray}
where
\begin{eqnarray}
 {\cal K}_A(p^2,q^2) :&=& d_G L(p^2,q^2,m_\Gp^2)+L(p^2,q^2,m_\Gs^2),
  \nonumber\\
 {\cal K}_B(p^2,q^2) :&=& d_G K(p^2,q^2,m_\Gp^2)-K(p^2,q^2,m_\Gs^2),
\end{eqnarray}
with
\begin{eqnarray}
 K(p^2,q^2,\Gm^2) &:=& {2 \over
  p^2+q^2+\Gm^2+\sqrt{(p^2+q^2+\Gm^2)^2-4p^2q^2}},
  \nonumber\\
 L(p^2,q^2,\Gm^2) :&=& {4p^2 q^2 \over
  [p^2+q^2+\Gm^2+\sqrt{(p^2+q^2+\Gm^2)^2-4p^2q^2}]^2}  .
\end{eqnarray}
Hence we obtain
\begin{eqnarray}
&& {\pd \over \pd J} \int {d^4k \over (2\Gp)^4}
 [G_\Gs(k) \GP_\Gs(k)+d_G G_\Gp(k) \GP_\Gp(k)]\Big|_{J^A=0}
\nonumber\\
&& =  {J \over 64\Gp^4}
\int  dp^2 \int  dq^2  \Biggr\{
\left({p^2+q^2 \over p^2 q^2} \right)  {\cal K}_A(p^2,q^2)
 - 2 {\cal K}_B(p^2,q^2) \Biggr\} + {\cal O}(J^3) .
\end{eqnarray}
\par
For the full fermion propagator of the form
${\cal S}(p)^{-1}=\slala p A(p^2)- B(p^2)$,
 ${\cal K}_A(p^2,q^2)$ and ${\cal K}_B(p^2,q^2)$
are integral kernels of the SD equation of the pure Yukawa model
($\Ga=0$) for $A$ and $B$, respectively,
see \cite{KTY90,Tanabashi91}.  The SD equation of the gauged Yukawa
model is given by
\begin{eqnarray}
 {\cal S}^{-1}(p) = S_0^{-1}(p) - g_Y \langle \Gs \rangle
 - \GS(p) - \GS_\Gs(p) - \GS_\Gp(p),
\end{eqnarray}
where
\begin{eqnarray}
 \GS(p) &:=& e^2 \int {d^4 k \over (2\Gp)^4}
              D_{\Gm \Gn}(k) \Gg_\Gm {\cal S}(p+k) \Gg_\Gn,
\nonumber\\
 \GS_\Gs(p) &:=& g_Y^2 \int {d^4 k \over (2\Gp)^4}
              G_{\Gs}(k) {\bf 1}{\cal S}(p+k) {\bf 1},
\nonumber\\
 \GS_\Gp(p) &:=& g_Y^2 \int {d^4 k \over (2\Gp)^4}
              G_{\Gp}(k) i\Gg_5 T_a {\cal S}(p+k) i\Gg_5 T_a .
\end{eqnarray}
This SD equation is decomposed into a pair of integral equations:
\begin{eqnarray}
 B(x) &=& g_Y \langle \Gs \rangle + {1 \over 16\Gp^2} \int_0^{\GL^2} dy
 {B(y) \over A^2(y)y+B^2(y)}
 [g_Y^2 {\cal K}_B(x,y)+(3+\Gx)e^2 K(x,y,0)],
\nonumber\\
 A(x) &=& 1 + {1 \over 32\Gp^2} \int_0^{\GL^2} dy
 {A(y) \over A^2(y)y+B^2(y)} {y \over x}
 [g_Y^2 {\cal K}_A(x,y)+2\Gx e^2 L(x,y,0)].
\end{eqnarray}
We should remark that the above SD equation for the fermion propagator
is obtained in the {\it quenched} approximation in the sense that the
photon propagator $D_{\Gm \Gn}(k)$ and the scalar ones $G_{\Gs}(k)$,
$G_{\Gp}(k)$ are all bare quantities.

\par
In what follows we consider the case of equal mass: $m_\Gs=m_\Gp=\Gm$.
\par
For the $SU(2)_L \times SU(2)_R$ symmetric Yukawa model,
${\cal K}_A(p^2,q^2) = 4L(p^2,q^2,\Gm^2)$,
${\cal K}_B(p^2,q^2) = 2K(p^2,q^2,\Gm^2)>0$, which implies the
attractive force. There is no problem in this case.
\par
For the $U(1)_L \times U(1)_R$ symmetric Yukawa model, however, a
specific cancellation occurs:  ${\cal K}_B(p^2,q^2)=0$,
while ${\cal K}_A(p^2,q^2) = 2L(p^2,q^2,\Gm^2)$.
This cancellation
of the attractive force for the $U(1)_L \times U(1)_R$ symmetric Yukawa
model  is pointed out \cite{KTY90,Tanabashi91} in the framework of the
Cornwall-Jackiw-Tomboulis (CJT) effective potential \cite{CJT74}
and the SD equation. When $\langle \Gs \rangle =0$, therefore, the
spontaneous  $\Gc$SB  does not occur within the
SD equation approach (to this order).
This type of peculiarity of U(1) Yukawa model is observed in the
analytical study of lattice Yukawa model \cite{EK92}.
\par
However,
the inversion method shows that, even in the U(1)-symmetric pure Yukawa
model,
$\Gc$SB does occur
$\langle \bar \Gy \Gy \rangle \not=0$
without acquiring the vacuum expectation value,
$\langle \Gs \rangle =0$,
at strong Yukawa coupling,
which is in accord with Monte Carlo result \cite{HN89}.
The origin of the difference between two approaches is seen as follows.
In this paper
the generating functional is calculated by closing the boson line of
the vacuum polarization diagram
($\GP_{\Gm\Gn}(k),\GP_\Gs(k),\GP_\Gp(k)$).
Alternatively, this is obtained by closing the electron line of the
fermion self-energy diagram ($\GS^J(p),\GS_\Gs^J(p),\GS_\Gp^J(p)$),
since
\begin{eqnarray}
 && \int {d^4k \over (2\Gp)^4}
 [D^{\Gm \Gn}(k) e^2 \GP_{\Gm \Gn}(k)
 + G_\Gs(k) g_Y^2 \GP_\Gs(k)
+  d_G G_\Gp(k) g_Y^2 \GP_\Gp(k) ]
 \nonumber\\
&&=
\int {d^4p \over (2\Gp)^4}
\tr\{S_J(p)[\GS^J(p)+\GS_\Gs^J(p)+\GS_\Gp^J(p)]\},
\end{eqnarray}
where
\begin{eqnarray}
 \GS^J(p) &:=& e^2 \int {d^4 k \over (2\Gp)^4}
              D_{\Gm \Gn}(k) \Gg_\Gm S_J(p+k) \Gg_\Gn,
\nonumber\\
 \GS_\Gs^J(p) &:=& g_Y^2 \int {d^4 k \over (2\Gp)^4}
              G_{\Gs}(k) {\bf 1}S_J(p+k) {\bf 1},
\nonumber\\
 \GS_\Gp^J(p) &:=& g_Y^2 \int {d^4 k \over (2\Gp)^4}
              G_{\Gp}(k) i\Gg_5 T_a S_J(p+k) i\Gg_5 T_a .
\end{eqnarray}
In this case, the explicit gauge dependence remains in the final
result and hence the gauge independence is lost even in the inversion
method as shown in pure QED \cite{UKF90},
which reflects the explicit gauge-parameter dependence of the SD
equation.
Such an inconvenience appears in the Yukawa model, too.
If we perform the calculation in the latter way,
the contribution to the fermion mass function $B(p^2)$ from the fermion
self-energy $\GS_\Gs^J(p)$ and $\GS_\Gp^J(p)$ cancels each other in the
$U(1)$ symmetric case for $m_\Gs=m_\Gp$, as observed in the SD equation
approach \cite{KTY90,Tanabashi91}.
To recover the result of the inversion method, the vacuum polarization
effect for the photon propagator and the scalar ones should be included
in the SD equation for the fermion propagator, as done in the gauged NJL
model.

\par
For the $Z(2)_L \times Z(2)_R$ symmetric Yukawa model,
${\cal K}_A(p^2,q^2) = L(p^2,q^2,\Gm^2)>0$,
${\cal K}_B(p^2,q^2) = -K(p^2,q^2,\Gm^2)<0$, which corresponds to the
repulsive force.  However the dynamical symmetry breaking in the above
sense occurs even in this case, which is consistent with the lattice
result.
\par
We notice
\begin{eqnarray}
&& {\pd \over \pd J} \int {d^4k \over (2\Gp)^4}
 [G_\Gs(k) \GP_\Gs(k)+d_G G_\Gp(k) \GP_\Gp(k)]\Big|_{J^A=0}
\nonumber\\
&=& {J \over 64\Gp^4}
\int^{\GL_f^2}  dp^2 \int^{\GL_f^2}  dq^2
\Biggr[ -(d_G+1){p^2+q^2 \over p^2 q^2}
\nonumber\\
&&
+{(d_G+1)(p^4+q^4)+4p^2 q^2+(d_G+1)\Gm^2(p^2+q^2) \over p^2 q^2}
K(p^2,q^2,\Gm^2)  \Biggr] + {\cal O}(J^3),
\label{scalarvp}
\end{eqnarray}
where we have used
\begin{eqnarray}
L(p^2,q^2,\Gm^2) = (p^2+q^2+\Gm^2)K(p^2,q^2,\Gm^2)-1.
\end{eqnarray}
To perform the above integration is not so easy.
To estimate the above quantity, therefore, we use the upper bound
\footnote{Validity of this approximation will be discussed later.}
for $K(p^2,q^2,\Gm^2)$:
\begin{eqnarray}
K(p^2,q^2,\Gm^2) \le {\Gq(p^2-q^2) \over p^2+\Gm^2}
+ {\Gq(q^2-p^2) \over q^2+\Gm^2},
\end{eqnarray}
where the equality holds when $\Gm=0$.
\footnote{Moreover, $K(p^2,q^2,\Gm^2)$ coincides with the right hand side
on the edge $p^2=0$ or $q^2=0$.}
Using the upper bound, we obtain
\footnote{We introduce the infrared cutoff $\Ge$ and take the limit
$\Ge \rightarrow 0$ after performing the integration.}
\begin{eqnarray}
\eq{scalarvp}
\le  J {(7-d_G)\GL_f^2 \over 64\Gp^4}
\left[1-{\Gm^2 \over \GL_f^2}
\ln \left(1+ {\GL_f^2 \over \Gm^2} \right) \right]
+ {\cal O}(J^3).
\end{eqnarray}
\par
In the gauged Yukawa model, therefore,
the nontrivial contribution from the vacuum polarization of the
scalar fields exists:
\begin{eqnarray}
&&{N N_f  \over 2}g_Y^2
 {\pd \over \pd J} \int  {d^4k \over (2\Gp)^4}
 [G_\Gs(k) \GP_\Gs(k)+d_G G_\Gp(k) \GP_\Gp(k)]\Big|_{J^A=0}
\nonumber\\
&\cong& {N N_f \GL_f^2 \over 4\Gp^2}J g_Y^2
{(7-d_G) \over 32\Gp^2}
\left[1-{\Gm^2 \over \GL_f^2}
\ln \left(1+ {\GL_f^2 \over \Gm^2} \right) \right]
+ {\cal O}(J^3).
\end{eqnarray}
\par
Now we take into account the  contribution from the vacuum
expectation value of the scalar field which obeys the relation:
\begin{eqnarray}
  [-\pd^2 +m_\Gs^2] \langle \Gs \rangle
  + \Gl \langle (\Gs^3+\Gs \Gp_a^2) \rangle
  = g_Y \langle \bar \Gy \Gy  \rangle.
\label{eqofmotion}
\end{eqnarray}
\par
When $\Gl=0$, we take $m_\Gs^2>0$ and \eq{eqofmotion} reduces to
\begin{eqnarray}
\langle \Gs \rangle
  = {g_Y \over m_\Gs^2} \langle \bar \Gy \Gy  \rangle ,
\end{eqnarray}
which yields
\begin{eqnarray}
 g_Y
 \langle \Gs \rangle
 {\pd \over \pd J} \langle \bar \Gy \Gy \rangle
&=& {1 \over 2} {g_Y^2 \over m_\Gs^2}
 {\pd \over \pd J} \langle \bar \Gy \Gy \rangle^2
\nonumber\\
&=&  {g_Y^2 \over m_\Gs^2}
     \left({N N_f\GL_f^2 \over 4\Gp^2}\right)^2
     \left[ J + 4{J^3 \over \GL_f^2}
     \ln \left({\GL_f^2 \over J^2}-1 \right) \right] +
{\cal O}(J^5).
\end{eqnarray}
Therefore the order parameter reads
\begin{eqnarray}
\Gf(J) &=&  {N N_f\GL_f^2 \over 4\Gp^2} J
\Biggr[ 1
+ {\Ga\over \Ga_c}
\nonumber\\
&&+ G_Y \left\{ 1+ {(7-d_G) \over 8N N_f}{m_\Gs^2 \over \GL_f^2}
- {(7-d_G) \over 8N N_f}\left({m_\Gs^2 \over \GL_f^2}\right)^2
\ln \left(1+ {\GL_f^2 \over m_\Gs^2} \right) \right\}
\Biggr]
\nonumber\\
&&- {N N_f \over 4\Gp^2 \Gh}
J^3 \left[4\Gh G_Y \ln {J^2 \over \GL_f^2}
 + {\Ga\over \Ga_c} \ln^2 {J^2 \over \GL_p^2} \right]
+ {\cal O}(J^3 \ln J^2),
\end{eqnarray}
where
\begin{eqnarray}
G_Y := {N N_f \over 4\Gp^2}{\GL_f^2 \over m_\Gs^2}g_Y^2 .
\end{eqnarray}

The inversion is performed in the same way as \eq{inversion} by
identifying $G_Y$ with $g$:
\begin{eqnarray}
J(\Gf) &=&  {4\Gp^2 \over N N_f\GL_f^2} \Gf
\Biggr[ 1
- {\Ga\over \Ga_c}
\nonumber\\
&-& G_Y
\left\{ 1+ {(7-d_G) \over 8N N_f}{m_\Gs^2 \over \GL_f^2}
- {(7-d_G) \over 8N N_f}\left({m_\Gs^2 \over \GL_f^2}\right)^2
\ln \left(1+ {\GL_f^2 \over m_\Gs^2} \right) \right\}
\Biggr]
\nonumber\\
&+&  {64 \Gp^6 \over \Gh N^3 N_f^3 \GL_f^8}
\Gf^3 \left[4\Gh G_Y \ln \left({16\Gp^4 \over N^2 N_f^2 \GL_f^6} \Gf^2
\right) + {\Ga\over \Ga_c} \ln ^2 \left({16\Gp^4 \over N^2 N_f^2\GL_f^4
\GL_p^2} \Gf^2 \right)   \right]
\nonumber\\
&+& {\cal O}(\Gf^3 \ln \Gf^2).
\end{eqnarray}

As a result, the critical Yukawa coupling $g_Y^c$ of the gauged Yukawa
model is obtained as
\begin{eqnarray}
  g_Y^c   =  {2\Gp  \over \sqrt{N N_f}}{m_\Gs \over \GL_f}
  \left(1 - {\Ga\over \Ga_c}\right)^{{1 \over 2}}
  \left[ 1+ {7-d_G \over 8N N_f}{m_\Gs^2 \over \GL_f^2}
- {7-d_G \over 8N N_f}\left({m_\Gs^2 \over \GL_f^2}\right)^2
\ln \left(1+ {\GL_f^2 \over m_\Gs^2} \right) \right]^{-{1 \over 2}}.
\end{eqnarray}
\par
Next we consider the phase in which $\langle \Gs \rangle =0$.
By neglecting the contribution from the tadpole term
$g_Y \langle \Gs \rangle \pd  \langle \bar \Gy \Gy \rangle /\pd J$,
the critical Yukawa coupling $g_Y^c{}'$ in such a phase is given by
\begin{eqnarray}
  g_Y^c{}'   =  \Gp \left({32 \over 7-d_G}\right)^{{1 \over 2}}
  \left(1 - {\Ga\over \Ga_c}\right)^{{1 \over 2}}
  \left[1 - {m_\Gs^2 \over \GL_f^2}
\ln \left(1+ {\GL_f^2 \over m_\Gs^2} \right) \right]^{-{1 \over 2}}.
\end{eqnarray}
We note that $g_Y^c{}'>g_Y^c$.
 From the above result,
$\langle \Gs \rangle \not=0$ and $\langle \bar \Gy \Gy \rangle \not=0$
for $g_Y>g_Y^c{}$.
In the region $g_Y>g_Y^c{}'$, both
$\langle \Gs \rangle =0$ and $\langle \bar \Gy \Gy \rangle \not=0$
hold simultaneously.
This shows that the chiral symmetry is dynamically
broken by the strong Yukawa coupling $g_Y>g_Y^c{}'$,  rather than the
nonvanishing vacuum expectation value of the scalar field due to the
double well potential.
Existence of such a phase is consistent with the result of
lattice Yukawa model \cite{HN89,EK92,Shen}.
\par
The estimation of the critical coupling
 $g_Y^c{}'=2\sqrt{2}\Gp$ for
$SU(2)$ symmetric pure Yukawa model
is not so bad in the light of
the numerical calculation \cite{KTY90} using the exact kernel
$K(p^2,q^2,m^2)$ of the quenched approximation in the sense stated
above.
The critical coupling $g_Y^c{}'$ will move into stronger Yukawa
coupling region, as $d_G$ increases.
For $SU(N)_L \times SU(N)_R$ symmetric Yukawa model,
$d_G=N^2-1$. Therefore our estimate must be improved for
$N \ge 3$.

\par
Finally we briefly discuss the case of $\Gl\not=0$.
In this case, the multi-point expectation value in
\eq{eqofmotion} will be truncated as
\begin{eqnarray}
 m_\Gs^2 \langle \Gs \rangle
  + 3\Gl  \langle  \Gs \rangle   \langle  \Gs^2 \rangle
  + \Gl  \langle  \Gs \rangle    \langle  \Gp_a^2 \rangle
  \cong g_Y \langle \bar \Gy \Gy  \rangle,
\end{eqnarray}
which yields
\begin{eqnarray}
  \langle \Gs \rangle
  \cong {g_Y \langle \bar \Gy \Gy  \rangle \over  m^2(\Gl)},
  \ m^2(\Gl)
  := m_\Gs^2+ \Gl (3\langle  \Gs^2 \rangle +  \langle  \Gp_a^2 \rangle ).
\end{eqnarray}
Therefore the order parameter reads
\begin{eqnarray}
\Gf(J) &=&  {N N_f\GL_f^2 \over 4\Gp^2} J
\Biggr[ 1+ G_Y^\Gl
\Big\{ 1+ {(7-d_G) \over 8N N_f}{m^2(\Gl) \over \GL_f^2}
\nonumber\\
&&- {(7-d_G) \over 8N N_f}\left({m^2(\Gl) \over \GL_f^2}\right)
\left({m_\Gs^2 \over \GL_f^2}\right)
\ln \left(1+ {\GL_f^2 \over m_\Gs^2} \right) \Big\} + {\Ga\over \Ga_c}
\Biggr]
\nonumber\\
&&- {N N_f \over 4\Gp^2 \Gh}
J^3 \left[4\Gh G_Y^\Gl \ln {J^2 \over \GL_f^2}
 + {\Ga\over \Ga_c} \ln^2 {J^2 \over \GL_p^2} \right]
+ {\cal O}(J^3 \ln J^2),
\end{eqnarray}
where
\begin{eqnarray}
G_Y^\Gl := {N N_f \over 4\Gp^2}{\GL_f^2 \over m^2(\Gl)}g_Y^2 .
\end{eqnarray}
Here the expectation value $\langle  \Gs^2 \rangle$ is calculated
at $\Gl=0$ for simplicity as follows:
\begin{eqnarray}
\langle  \Gs(x)^2 \rangle
= \int^{\GL_\Gs} {d^4k \over (2\Gp)^4} {1 \over k^2+m_\Gs^2}
= {\GL_\Gs^2 \over 16\Gp^2} \left[1-{m_\Gs^2 \over \GL_\Gs^2}
\ln \left(1+{\GL_\Gs^2 \over m_\Gs^2}\right) \right].
\end{eqnarray}
Then we obtain
\begin{eqnarray}
{4\Gp^2 m^2(\Gl) \over \GL_f^2}
= {4\Gp^2 m_\Gs^2 \over \GL_f^2}
  + \Gl {3\Gh_\Gs+\Gh_\Gp \over 4}
 - 3{m_\Gs^2 \over 4\GL_f^2} \ln \left(1+ {\GL_f^2 \over m_\Gs^2} \right)
 - {m_\Gp^2 \over 4\GL_f^2} \ln \left(1+ {\GL_f^2 \over m_\Gp^2} \right).
\end{eqnarray}
Neglecting the logarithmic term of the order
${\cal O}\left({m_\Gs^2 \over \GL_\Gs^2}
\ln \left(1+{\GL_\Gs^2 \over m_\Gs^2}\right) \right)$,
the critical line of the gauged Yukawa model is obtained as
\begin{eqnarray}
  g_Y^2   \cong  {1 \over N N_f}
  \left[ {4\Gp^2 m_\Gs^2 \over \GL_f^2}
  + \Gl {3\Gh_\Gs+\Gh_\Gp \over 4}
  \right]
  \left(1 - {\Ga\over \Ga_c}\right),
\end{eqnarray}
where
\begin{eqnarray}
\Gh_\Gs := {\GL_\Gs^2 \over \GL_f^2},
\Gh_\Gp := {\GL_\Gp^2 \over \GL_f^2}.
\end{eqnarray}

\section{Conclusion and Discussion}
In this paper
we have shown the existence of the second order  chiral phase transition
in the gauged NJL and the gauged Yukawa models by use of the inversion
method. By calculating the gauge-invariant chiral condensate
$\langle \bar \Gy \Gy \rangle$ in the inversion
method to the lowest order, we have derived the critical line
separating   the spontaneous $\Gc$SB strong coupling phase from the
chiral symmetric weak coupling one.
The chiral order parameter $\Gf = \langle \bar \Gy \Gy \rangle$ has been
shown to take  the mean-field critical exponent near the whole critical
line, together with possible logarithmic correction to the mean-field
prediction.
\par
The above results obtained in the inversion method have been compared
with  the previous results obtained from the SD equation for the
fermion propagator in the special (Landau) gauge \cite{Kondo92p}.\
Drawback of the quenched SD equation in the $U(1)_L \times U(1)_R$
symmetric Yukawa model is removed in the inversion method.
This enables us to realize the top quark condensation through the
strong Yukawa coupling \cite{KTYprep}.
We emphasize that all the above results obtained
in the inversion method are gauge-parameter independent, which show
validity of the SD equation approach.
\par
Finally, in order to derive $N_f$-dependence of the critical
coupling $\Ga_c$ and to reproduce the critical line in quite large
coupling region in the gauged NJL model,
we will need to evaluate the next to
leading order and higher order in the inversion method.

\section{Acknowledgments}
The author would like to thank Koichi Yamawaki for interest in this
work and Takeshi Inagaki for critical comments and sending him useful
informations on the inversion method.

\end{document}